\documentclass[12p]{iopart}
\usepackage{import}
\usepackage[dvipsnames, table]{xcolor}
\usepackage{graphicx}
\usepackage{subcaption}
\usepackage{url}
\usepackage{cite}

\usepackage{hyperref}
\usepackage{comment}
\usepackage{adjustbox}
\makeatletter
\providecommand{\doi}[1]{%
  \begingroup
    \let\bibinfo\@secondoftwo
    \urlstyle{rm}%
    \href{http://dx.doi.org/#1}{%
      doi:\discretionary{}{}{}%
      \nolinkurl{#1}%
    }%
  \endgroup
}
\makeatother
\usepackage[nameinlink,noabbrev,capitalize]{cleveref}

\renewcommand{\cref}{\Cref}

\let\oldexp\exp
\renewcommand{\exp}[1]{\oldexp{\left( {#1} \right)}}

\begin{document}

\title{HoloTile Light Engine: New Digital Holographic Modalities and Applications}
\author{Jesper Gl\"uckstad, Andreas Erik Gejl Madsen}

\address{SDU Centre for Photonics Engineering, University of Southern Denmark, Campusvej 55, 5230 Odense-M}
\ead{jegl@sdu.dk, gejl@mci.sdu.dk}

\begin{abstract}
	HoloTile is a patented computer generated holography approach with the aim of reducing the speckle noise caused by the overlap of the non-trivial physical extent of the point spread function in Fourier holographic systems from adjacent frequency components. By combining tiling of phase-only of rapidly generated sub-holograms with a PSF-shaping phase profile, each frequency component - or output ``pixel" - in the Fourier domain is shaped to a desired non-overlapping profile. In this paper, we show the high-resolution, speckle-reduced reconstructions that can be achieved with HoloTile, as well as present new HoloTile modalities, including an expanded list of PSF options with new key properties. In addition, we discuss numerous applications for which HoloTile, its rapid hologram generation, and the new PSF options may be an ideal fit, including optical trapping and manipulation of particles, volumetric additive printing, information transfer and quantum communication.
\end{abstract}


\maketitle
\ioptwocol

\begin{figure*}
	\centering
	\begin{subfigure}{0.3\textwidth}
	\includegraphics[width=\textwidth]{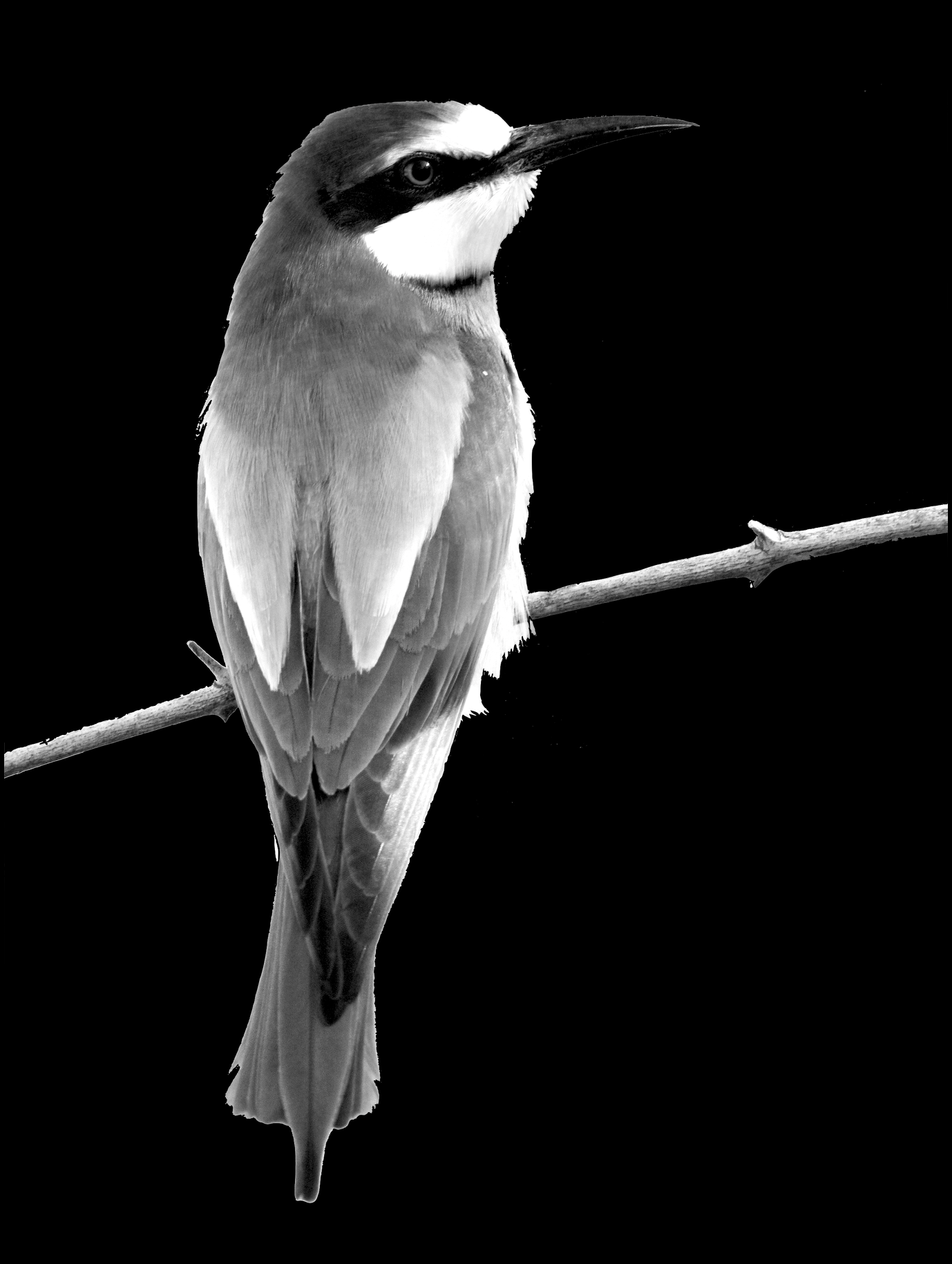}
	\caption{\small Reconstruction target}
	\label{fig:humming-target}
	\end{subfigure}
	\begin{subfigure}{0.3\textwidth}
	\includegraphics[trim={0 0.3cm 0 0.19cm}, clip, width=\textwidth]{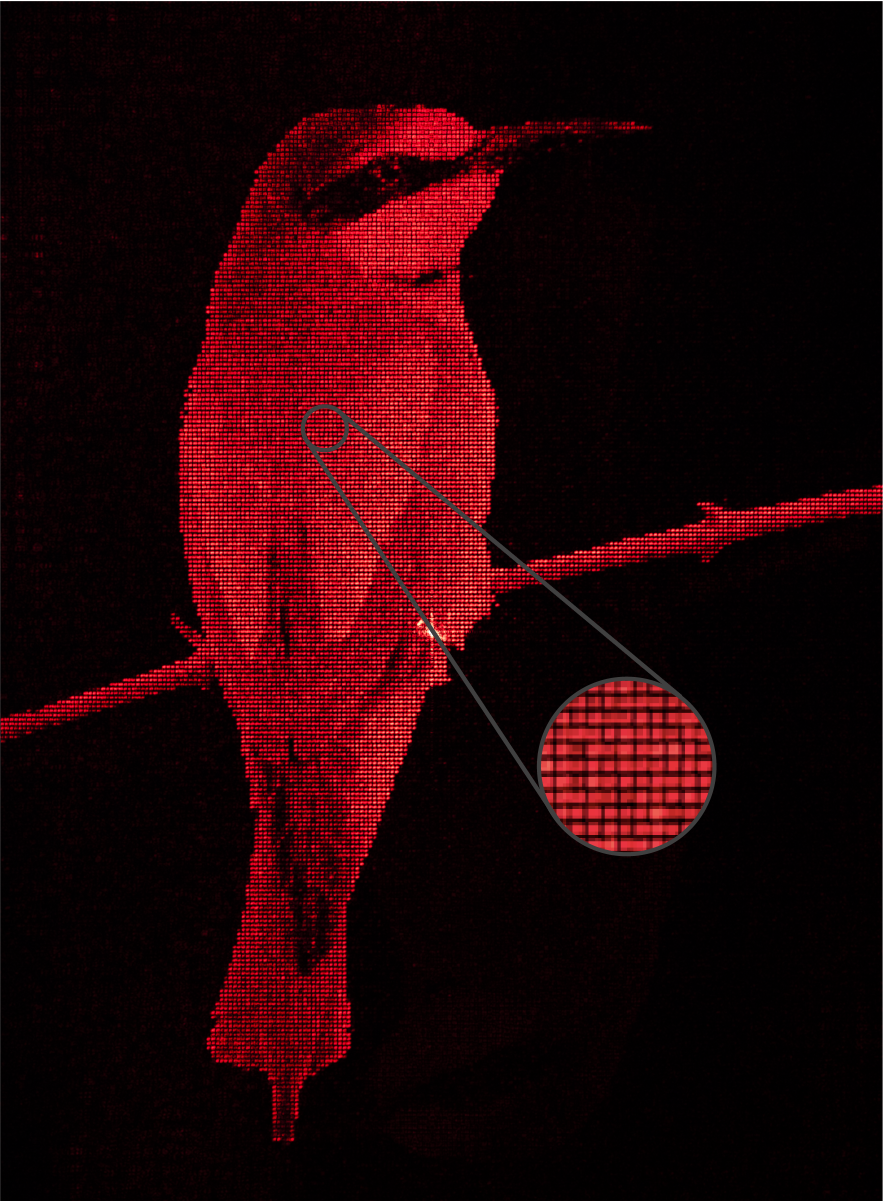}
	\caption{\small Exp. (lensed) reconstruction}
	\label{fig:humming-recon-lensed}
	\end{subfigure}
	\begin{subfigure}{0.3\textwidth}
	\includegraphics[trim={0 0cm 0 0.05cm},clip, width=\textwidth]{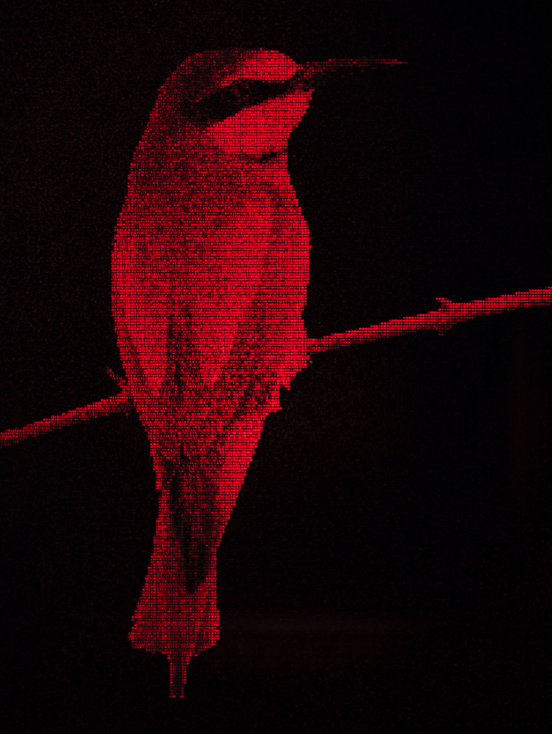}
	\caption{\small Exp. (lensless) reconstruction}
	\label{fig:humming-recon-lensless}
	\end{subfigure}
	\begin{subfigure}{0.4\textwidth}
		\includegraphics[trim={22cm 18.4cm 22cm 19.7cm},clip,width=\textwidth]{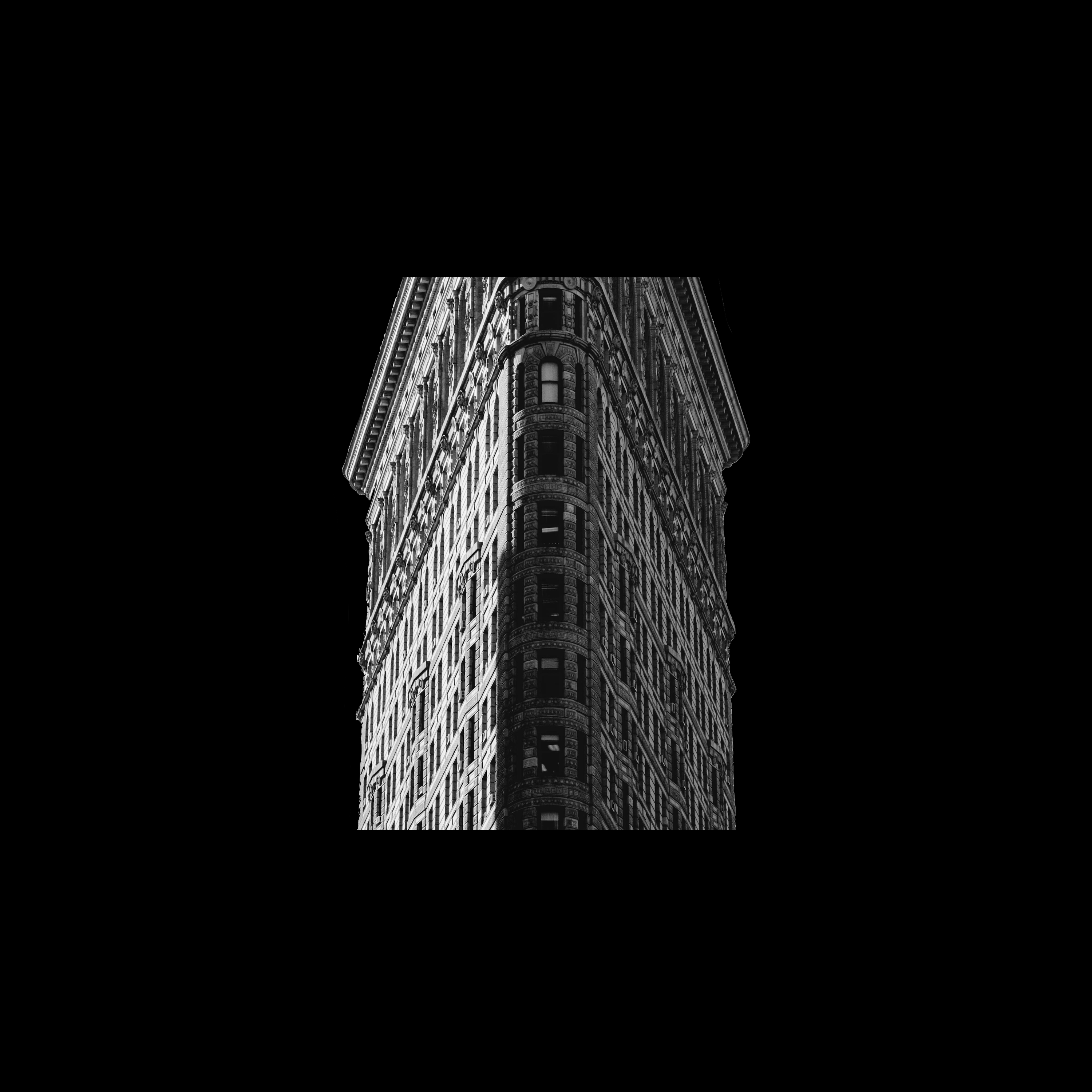}
		\caption{\small Reconstruction target}
		\label{fig:flatiron}
	\end{subfigure}
	\begin{subfigure}{0.4\textwidth}
		\includegraphics[width=\textwidth]{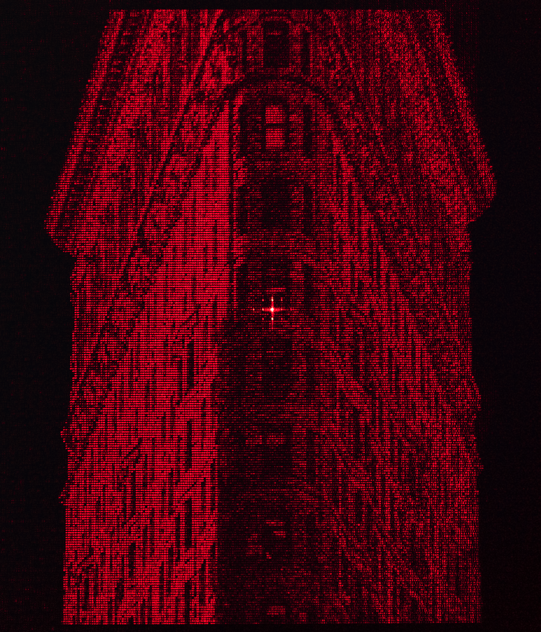}
		\caption{\small Experimental (lensed) reconstruction}
		\label{fig:flatiron-exp}
	\end{subfigure}
	\begin{subfigure}{0.4\textwidth}
		\includegraphics[trim={15.5cm 21.3cm 15.5cm 21.5cm},clip,width=\textwidth]{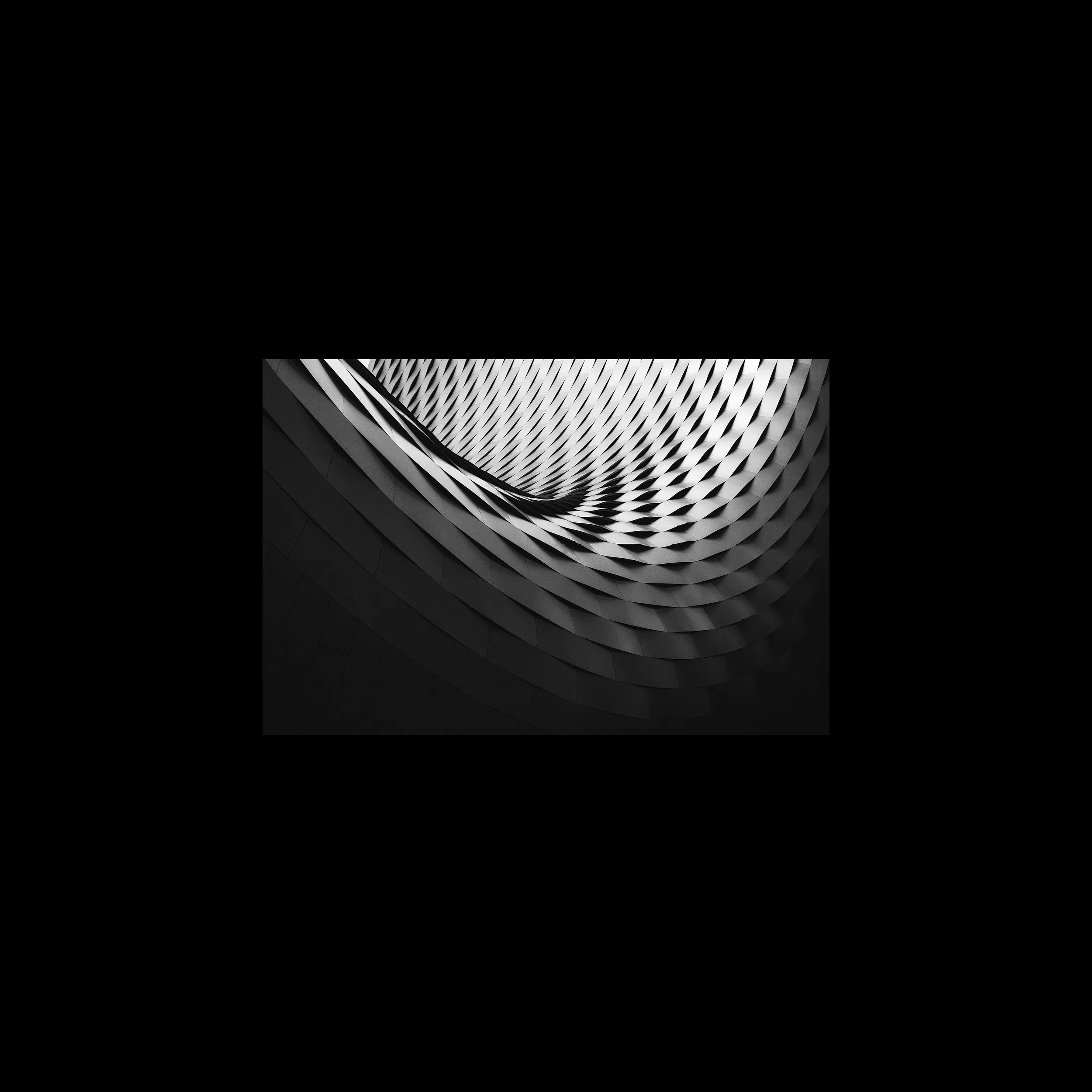}
		\caption{\small Reconstruction target}
		\label{fig:building-target}
	\end{subfigure}
	\begin{subfigure}{.4\textwidth}
		\includegraphics[width=\textwidth]{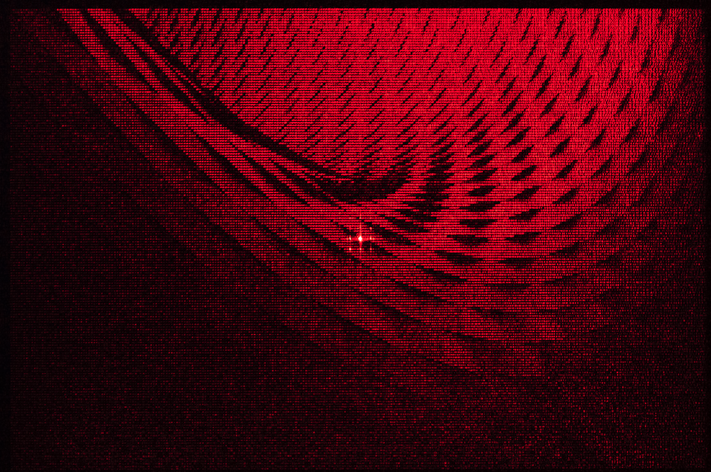}
		\caption{\small Experimental (lensed) reconstruction}
		\label{fig:building-1}
	\end{subfigure}
	\caption{Targets and experimental HoloTile reconstructions at low tile-numbers ($N_t=4$) as captured using the setup in \cref{fig:optical-setup}.}
\end{figure*}

\section{Background}
Modern science heavily relies on the capacity to dynamically manipulate both coherent and semi-coherent light in two and three dimensions, which has opened new doors for numerous studies and practical applications. Such flexibility in light sculpting has proven its utility as a universal tool that enhances control over light-matter interaction in a volumetric manner. This has spurred innovation in fields such as neuroscience\cite{papagiakoumou_scanless_2010,papagiakoumou_optical_2013}, microbiology, microscopic optical manipulation \cite{grier_revolution_2003,rodrigo_real-time_2004,rodrigo_actuation_2005}, non-invasive cell sorting techniques \cite{banas_using_2012,banas_matched-filtering_2012-1}, materials processing \cite{olsen_multibeam_2009}, and microfabrication \cite{kawata_finer_2001,galajda_complex_2001}, to mention a few. Additionally, it has lent itself to applications involving controlled photo-stimulation \cite{papagiakoumou_scanless_2010,go_simultaneous_2012}, cell surgery \cite{ando_optical_2008}, and advanced light microscopy \cite{hell_breaking_1994,smith_programmable_2000}.

Due to the diversity of applications and methodologies involved, making a choice on the most effective light sculpting technique for a specific purpose becomes a critical task. For instance, the need for strong gradient forces for position stability \cite{ashkin_observation_1985} or for beams that can act upon large areas \cite{ulriksen_independent_2008} can dictate the use of optical micro-manipulation in certain experiments. Static beams might be sufficient for microfabrication but it requires intense, highly localized light for triggering nonlinear processes with high precision.

In addition to application-specific needs, other pragmatic factors like photon efficiency, budget, and equipment size are also key considerations. In the context of requirements, photon efficiency is a critical yet straightforward factor, making it a preferred attribute in laser applications involving light sculpting. This underlines why phase-only modulation techniques are typically preferred over more straightforward absorbing amplitude methods. These techniques manipulate available light through interference and diffraction effects rather than diverting or blocking light to achieve a desired distribution. This is exemplified by Computer Generated Holography (CGH), a specific form of phase-only light sculpting that's increasingly adopted across diverse industrial and research applications.

\begin{figure*}
	\centering
	\includegraphics{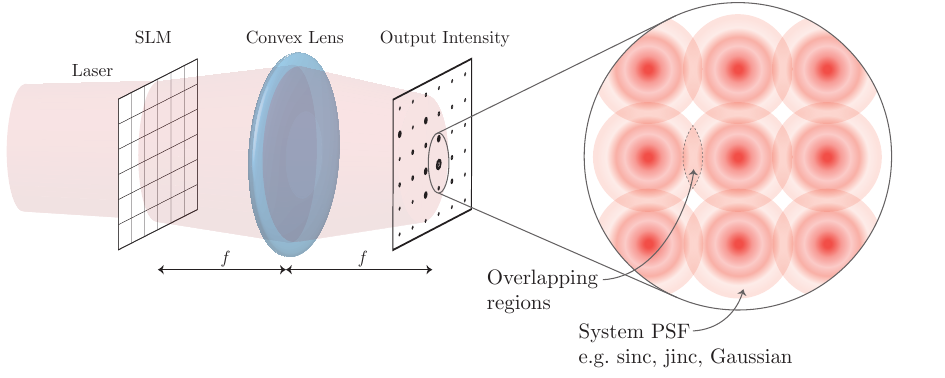}
	\caption{Illustration of a common problem of holographic Fourier engines. The PSF of the system optics in each spatial frequency coordinate may overlap in the reconstruction plane. When the phase profile in this plane is unconstrained, these overlaps may lead to unwanted interference i.e. speckle noise.}
	\label{fig:fourier-engine}
\end{figure*}

\begin{figure*}
	\centering
	\includegraphics[width=0.9\textwidth]{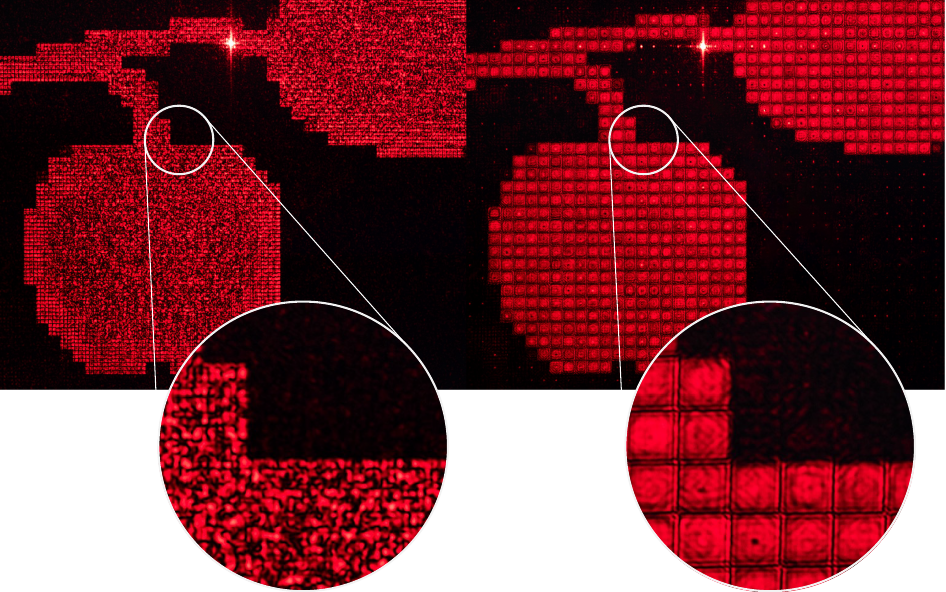}
	\caption{Comparison of homogeneity and output regularity between (Right) HoloTile and the (Left) Adaptive Weighted Gerchberg Saxton (AWGS) algorithm \cite{wu_adaptive_2021}. Reconstructions as captured in camera with phase modulation by the HoloEye GAEA SLM. The output ``pixel" structure encoded in the HoloTile holograms results in the target being defined on a regular grid. The HoloTile reconstruction shows limited cross-talk between output frequency components as opposed to the speckle noise seen in the AWGS reconstruction.}
	\label{fig:homogeneityComparison}
\end{figure*}

The advantages of possible 3-dimensional display projection and high photon efficiency are extremely appealing. However, the issue of speckle noise still cause significant problems for applications where homogeneity is key. Especially for applications that employ some version of two-photon stimulation, as the response is proportional to the square of light intensity, and not only intensity. With HoloTile \cite{gluckstad_holographic_2022,madsen_holotile_2022}, we have presented a patented method aimed at mitigating one source of the problematic speckle noise. Namely the unwanted interference between adjacent frequency components in Fourier holography caused by both the physical extent of the PSF of the system in the Fourier domain, as well as the undetermined phases in the reconstruction plane, as illustrated in \cref{fig:fourier-engine}. In the original paper, we describe a method combining sub-hologram tiling with clever PSF-shaping in order to avoid this nearest-neighbor interference in the spatial Fourier plane of CGH reconstruction as is demonstrated in \cref{fig:homogeneityComparison}. The results showed high contrast reconstructions with superior homogeneity and regularity in the reconstrctions, in addition to extremely fast computation (see \cref{tab:computationTime}) due to the constant PSF shaping phase (calculated once), and the lower-resolution sub-holograms.
\begin{figure*}
	\centering
	\includegraphics[width=\textwidth]{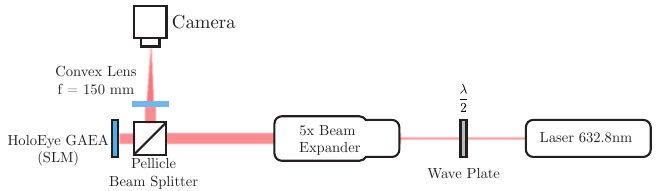}
	\caption{Illustration of the optical setup used for all experimental results in this paper. A helium-neon laser is shone through a half-wave plate and a $5\times$ beam expander onto a HoloEye GAEA\cite{noauthor_gaea-2_2022} SLM which modulates the incoming light. The reconstructions are captured by a Canon M6 Mark II camera.}
	\label{fig:optical-setup}
\end{figure*}
HoloTile (experimentally shown in \cref{fig:humming-recon-lensed,fig:flatiron-exp,fig:building-1}) aims to solve the challenge of rapid and speckle-free light sculpting without the need for time-averaging techniques - a challenge that exists in several fields of optics, biophotonics, additive manufacturing, display technology and others. The rapid nature of HoloTile hologram generation ($>100\times$ speed-improvement\cite{madsen_holotile_2022}) potentially makes the method attractive for a variety of applications including
\begin{itemize}
	 \item Ultrafast additive manufacturing
	\item Laser material processing in parallel
	\item Holographic volumetric bioprinting
	\item Rapid laser engraving, welding, machining
	\item Two-photon excitation in optogenetics
	\item Neurophotonics voltage imaging
	\item Multi-color and multi-plane diffraction
	\item Photon-efficient phase-only display technology
	\item Real-time adaptive optics embodiments including aberration correction
	\item Temporal focusing holography for parallel micron-depth slice-excitation
	\item Holographic light-sheet based microscopy
	\item Parallel super-resolution microscopy such as STED
	\item Phase-only holographic security and cryptology
	\item Digital quantum holography
\end{itemize}
In future research and development we aim to demonstrate some of these potential advantages of HoloTile in one or more of the above light diffraction applications. Our ultimate aim is to offer HoloTile as a stand-alone light engine that can be integrated with ease in both hardware and software in existing optics and photonics configurations for both industry and academia. 
To start, in this paper, we will introduce new modalities of the HoloTile method, as well as provide examples of applications in which the method could excel.

\begin{table}[]
\caption{Calculation times for both AWGS and HoloTile to create the holograms reconstructed in \cref{fig:homogeneityComparison}. Both are calculating $2160\times 2160$ pixel holograms, and are shown calculated on the CPU (13th Gen Intel(R) Core(TM) i9-13900HX)  and GPU (NVIDIA GeForce RTX 4080 Laptop).}
\label{tab:computationTime}
\begin{tabular}{lc}
\textbf{Algorithm}      & \multicolumn{1}{l}{\textbf{Calculation Time {[}ms{]}}} \\ \hline 
   
\textbf{AWGS (CPU)}     & 2923.26 \rule{0pt}{2.6ex} 								\\
\textbf{AWGS (GPU)}     & 149.64                                                 \\
\textbf{HoloTile (CPU)} & 14.22                                                  \\
\textbf{HoloTile (GPU)} & 8.15                                                  
\end{tabular}
\end{table}

\section{PSF Shaping Modalities}
While the square top-hat output pixel shape described in the original HoloTile paper \cite{madsen_holotile_2022} (\cref{fig:field-square}) has numerous uses, the process by which it is generated enables the straightforward implementation of virtually endless other pixel shapes to suit any particular application.
\subsection{Top-hat Disk Pixels}
By employing the same stationary phase method used for the generation of square output pixels \cite{dickey_laser_2014}, the phase function for converting an input Gaussian beam into a flat-top disk shape at the output can be expressed as \cite{dickey_gaussian_1996, romero_lossless_1996}:
\begin{equation}
\label{eq:disk-phase}
	\phi_\textrm{disk}(\xi) = \beta \frac{\sqrt{\pi}}{2} \int_0^\xi \sqrt{1 - \exp{(-\rho^2)}} \, \textrm{d}\rho
\end{equation}
where $\xi = \frac{\sqrt{2} r}{r_0}$, $\beta = \frac{2\sqrt{2\pi} r_0 y_0}{f \lambda}$,
$r$ is the radius from the optical axis in the hologram plane, $r_0$ is the $1/e^2$ radius of the incident Gaussian beam, $y_0$ is the radius of the desired disk shape, $f$ is the focal length of the Fourier transforming lens, and $\lambda$ denotes wavelength. An experimental disk-HoloTile reconstruction is shown in \cref{fig:field-disk}. While the method of stationary phase theoretically allows for the conversion of any single-mode beam into arbitrary output irradiance profiles, several known and analytical solutions are readily available in the existing literature.

\subsection{Ring Pixels}
To generate ring shaped output pixels, which commonly find uses in materials processing \cite{grunewald_influence_2021,bischoff_design_2019} and optical particle manipulation\cite{zhao_trapping_2020}, an adjustable axicon phase can be employed:
\begin{equation}
\label{eq:ring-phase}
	\phi_\textrm{ring}(\xi) = \beta \xi
\end{equation}
Where $\beta$ is defined as above. The radius of the output ring can be adjusted by simply manipulating $y_0$ in the definition of $\beta$. Experimental examples of ring-HoloTile are shown in \cref{fig:field-ring}. It is worth noting that due to the increased localization of the focused light, the contrast of the ring reconstructions is improved significantly.

\subsection{Line Pixels}
The square top-hat pixel generating phase can easily be repurposed to generate 1D ``line pixels" that can be manipulated in a scanning fashion. By further scaling either $\xi$ or $\eta$, the already scaled lateral coordinates of the hologram plane, by a \textit{definition factor} $\gamma$, the horizontal or vertical extent of the output pixels can be modulated. For example, the vertical coordinate may be scaled as $\eta = \frac{\sqrt{2}y}{\gamma r_0}$, such that as $\gamma$ increases, the vertical extent of the now rectangular output pixel decreases. By carefully selecting $\gamma$, the desired width of the line pixel can be chosen. The output reconstructions are extremely well-defined spatially with high contrast. Experimental reconstructions are shown in \cref{fig:field-line}.

\subsection{Multi-line Pixels}
Combining multiple PSFs allows for more complex output pixel shapes. For instance, to create ``cross" shaped pixels, two line shaping PSF phase profiles may be combined, with one rotated by $90^\circ$. The combination of two PSF phase profiles have here been implemented by spatially multiplexing on the SLM in a checkerboard fashion. Furthermore, the output pixel shapes can be rotated freely by simply rotating the PSF shaping phase around the SLM center normal. An example of a cross-shaped PSFs is demonstrated in the experimental reconstructions of \cref{fig:field-cross}.

\subsection{Complex Light Distributions}
\label{sec:hc}
\begin{figure*}
    \centering
    \begin{subfigure}{0.45\textwidth}
        \includegraphics[width=\textwidth]{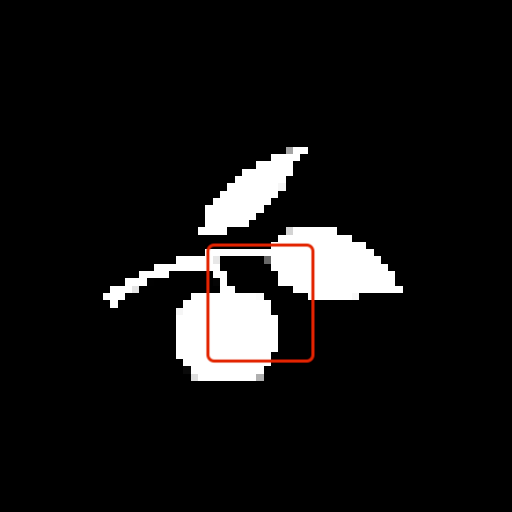}
        \caption{\small Reconstruction target}
        \label{fig:field-target}
    \end{subfigure}\\
    \begin{subfigure}{0.3\textwidth}
        \includegraphics[width=\textwidth]{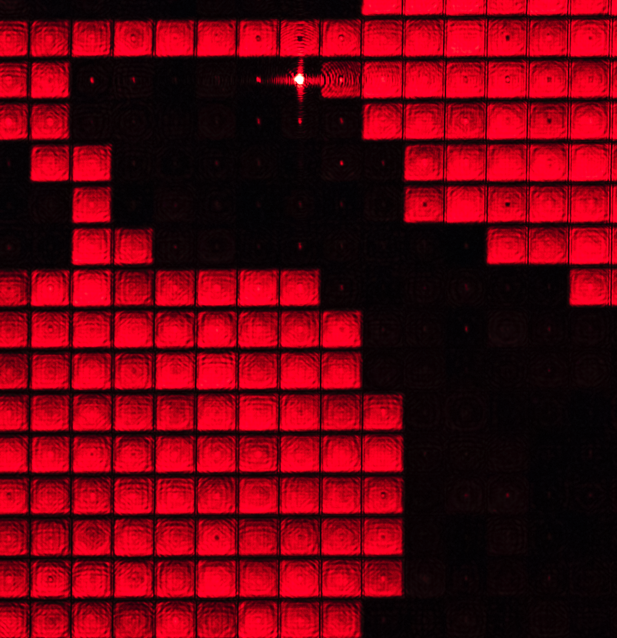}
        \caption{\small Square output pixels}
        \label{fig:field-square}
    \end{subfigure}
	\hfil
    \begin{subfigure}{0.3\textwidth}
        \includegraphics[width=\textwidth]{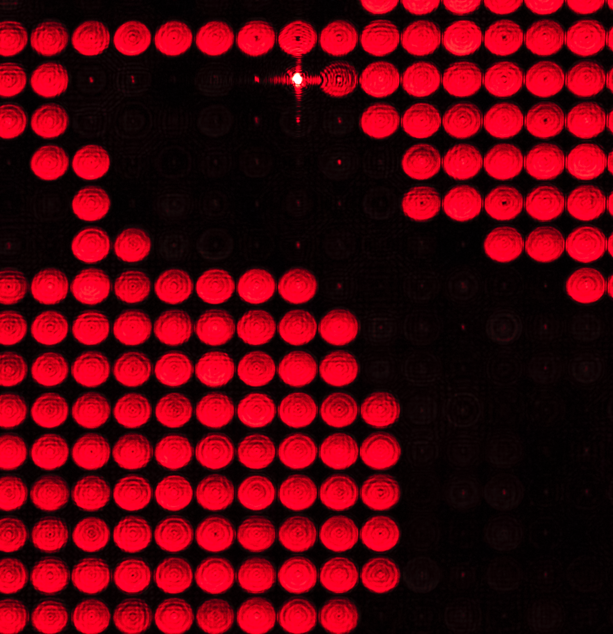}
        \caption{\small Disk output pixels}
        \label{fig:field-disk}
    \end{subfigure}
    \hfil
    \begin{subfigure}{0.3\textwidth}
        \includegraphics[width=\textwidth]{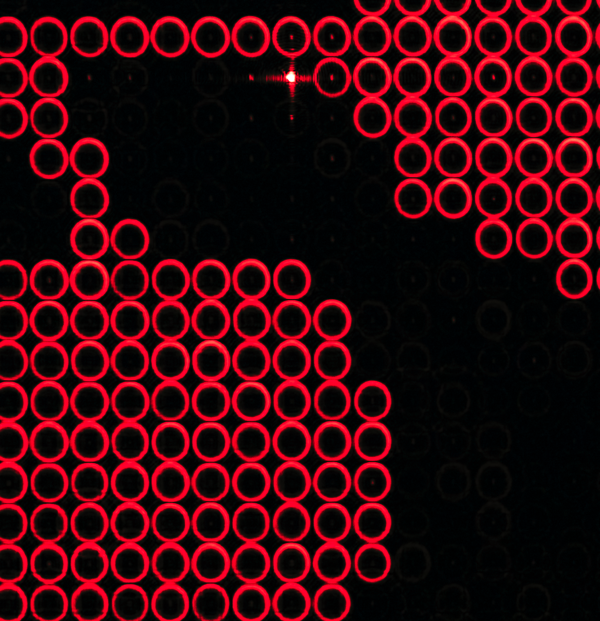}
        \caption{\small Ring output pixels}
        \label{fig:field-ring}
    \end{subfigure} \\
    \begin{subfigure}{0.3\textwidth}
        \includegraphics[width=\textwidth]{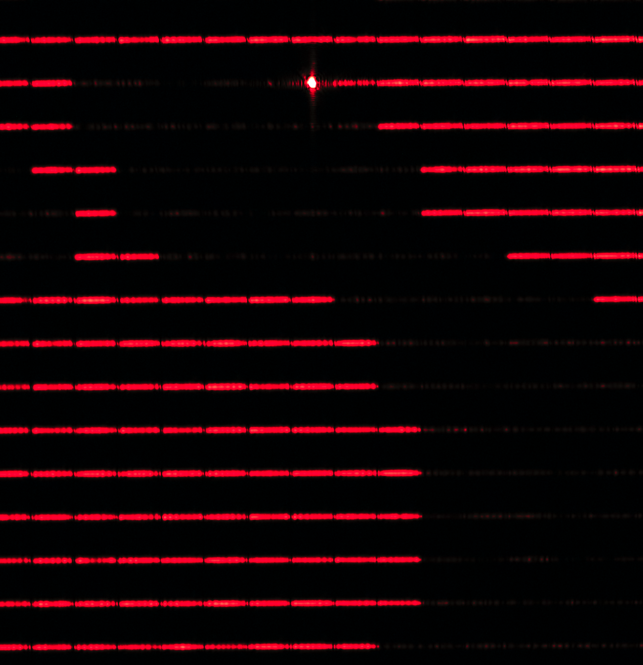}
        \caption{\small Line output pixels}
        \label{fig:field-line}
    \end{subfigure}
    \hfil
    \begin{subfigure}{0.3\textwidth}
        \includegraphics[width=\textwidth]{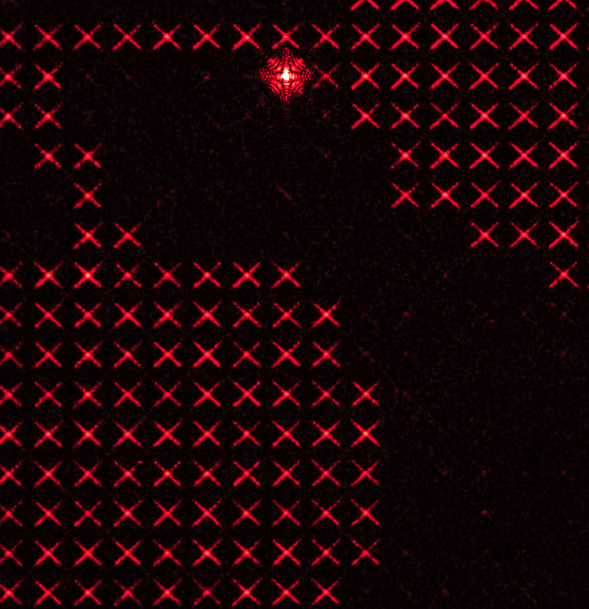}
        \caption{\small Cross output pixels}
        \label{fig:field-cross}
    \end{subfigure}
    \begin{subfigure}{0.3\textwidth}
        \includegraphics[width=\textwidth]{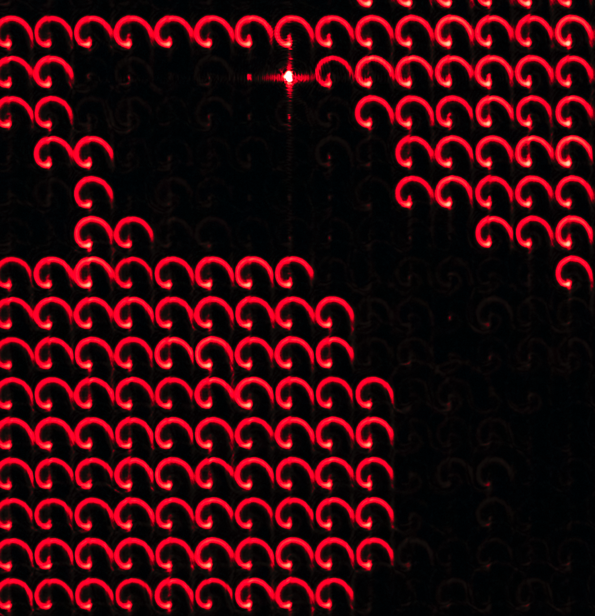}
        \caption{\small Helico-conical output pixels}
        \label{fig:field-hc}
    \end{subfigure}
    \caption{The PSF shaping capabilities of HoloTile shown experimentally. For identical targets, the output pixels can be shaped near arbitrarily. Experimental reconstructions are enlarged for clarity to the area of the red square in \cref{fig:field-target}.}
    \label{fig:field}
\end{figure*}

As the PSF shape is entirely independent of the sub-holograms, any engineered beam that can be coded on the SLM can be applied to HoloTile directly to facilitate fully controllable parallel processing. Light distributions such as optical vortices, conical, helical, and Airy beams, etc., find uses in optical trapping and manipulation of particles, super-resolution microscopy, high-capacity data transmission, 3D imaging, lithography and so on.
An example of such an engineered light distribution is the helico-conical beam \cite{alonzo_helico-conical_2005,daria_optical_2011,overton_phase_2005,engay_interferometric_2019}, which generates a spiral intensity pattern at the focal plane. The helico-conical generating phase profile is defined as the product of a helical and conical phase:
\begin{equation}
	\phi_{HC}(\xi) = \ell \theta  \left(K - \frac{\xi}{\sqrt{2}}\right)
\end{equation}
where $\ell$ is a scalar determining the number of $2\pi$ phase shifts across the azimuthal angle $\theta$ of the SLM, and $K$ is a constant that is either 1 or 0.
In \cref{fig:field-hc}, the experimental reconstruction of the helico-conical HoloTile modality is shown. The spiral patterns are contained well within each unit output pixel, with good contrast.

\section{Lens-less HoloTile}
Although the optical setup described in the original paper \cite{madsen_holotile_2022} is very simple, making use of a single convex lens as the Fourier transforming engine from the SLM plane to the reconstruction plane, HoloTile can be realized in a completely lens-less setup. A lens-less HoloTile configuration is particularly attractive for applications where cost and/or mechanical complexity must be kept to a minimum. The working principle of HoloTile requires a Fourier transform relationship between input and output planes in order to achieve proper separation of frequency components and PSF shaping. Therefore, a lens-less configuration could be constructed where the modulated beam is allowed to propagate until the far-field condition is met \cite{gluckstad_new_2023}. However, we have chosen to employ an additional quadratic lens phase profile that can be coded directly on the SLM in conjunction with the tiled sub-holograms and the PSF-shaping phase:
\begin{equation}
	\phi_{SLM} = \phi_{tiles} + \phi_{psf} + \phi_{lens}
\end{equation}
where
\begin{equation}
	\phi_{lens} = \frac{\pi}{\lambda f} r^2
\end{equation}
is the Fourier transforming lens phase defined by the desired focal length $f$.
The addition of the lens phase results in a reconstruction plane at a distance $f$ from the SLM (as opposed to $2f$ as is the case for the lensed configuration). This corresponds to a case where the physical Fourier transforming lens is placed immediately after the SLM. As such, a quadratic phase factor is introduced in the reconstruction plane \cite{goodman_introduction_2017}, however as only the intensity of the output pattern is of importance, this is of little concern.

With the lens phase in place, the HoloTile hologram generation can proceed as normal, with optimization of the tiled sub-hologram and design of the PSF-shaping phase.
In \cref{fig:humming-recon-lensless}, the experimental reconstruction of the hummingbird target from the lensless configuration is shown. Here, the focal length of the quadratic lens phase is set to $f=150 \textrm{ mm}$.

\section{Axial Propagation Characteristics}
Along with lateral control of the beamshapers, the axial propagation can also be made more predictable or tailored for specific applications. In traditional holographic generation methods, the reconstructions are only controllable in one or more reconstruction planes. As we move away from any of these, each point defocuses according to the NA of the Fourier transforming lens. Due to the unconstrained phases in the reconstruction planes, the field immediately interferes with itself to form an indeterminate noise pattern.

\begin{figure*}
    \centering
    \subcaptionbox{Sim, square.}{
    	\includegraphics[height=7.2cm]{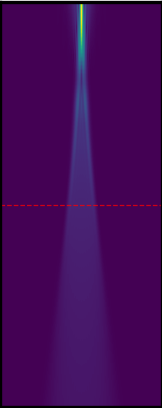}
    }
    \subcaptionbox{Experimentally captured reconstructions using square beamshaper.}{
    	\includegraphics[height=7.2cm]{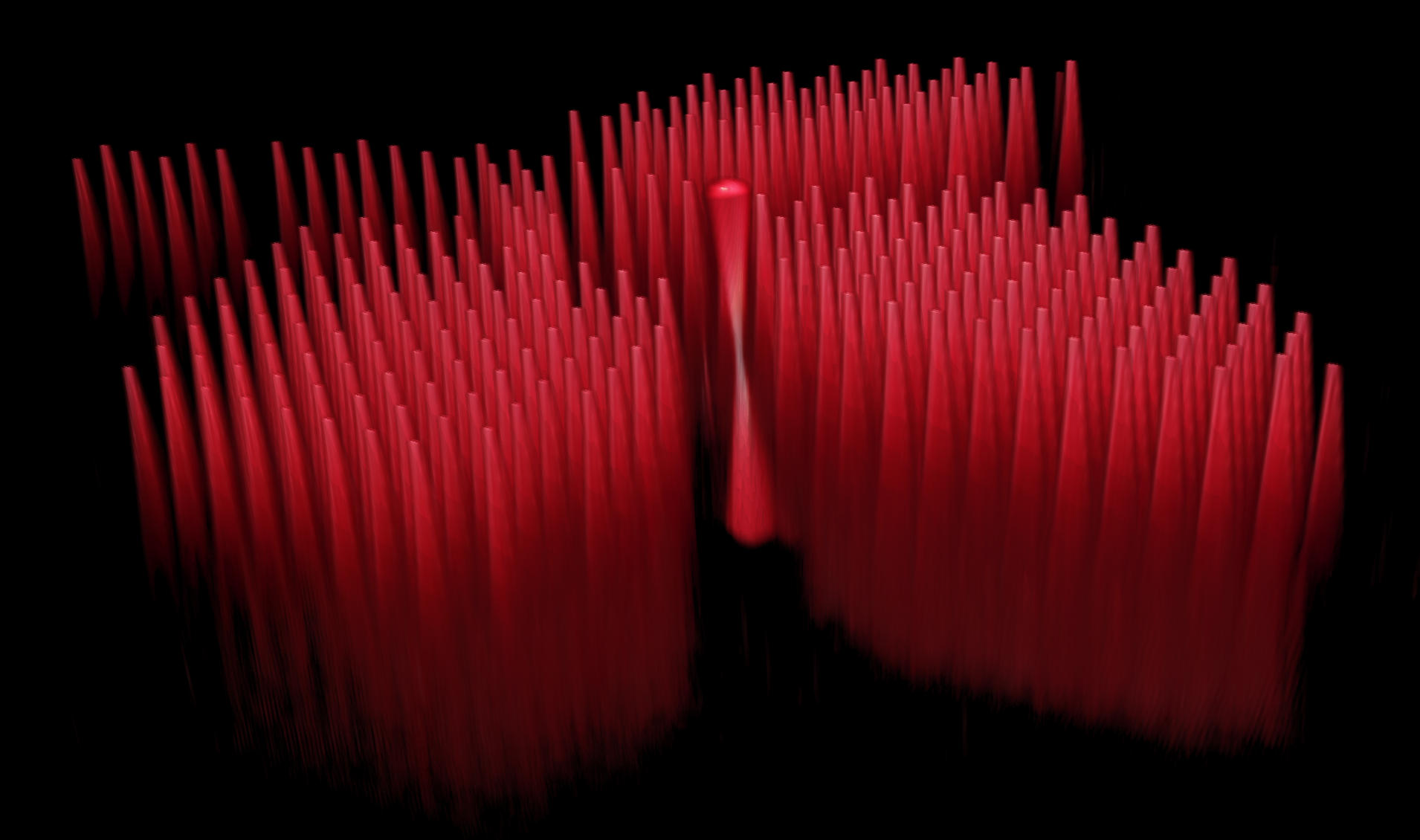}
    }
    \\
    \subcaptionbox{Sim, ring.}{
    	\includegraphics[height=7.1cm]{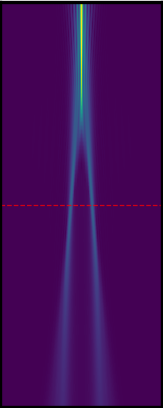}
    }
    \subcaptionbox{Experimentally captured reconstructions using ring beamshaper.}{
    	\includegraphics[height=7.1cm]{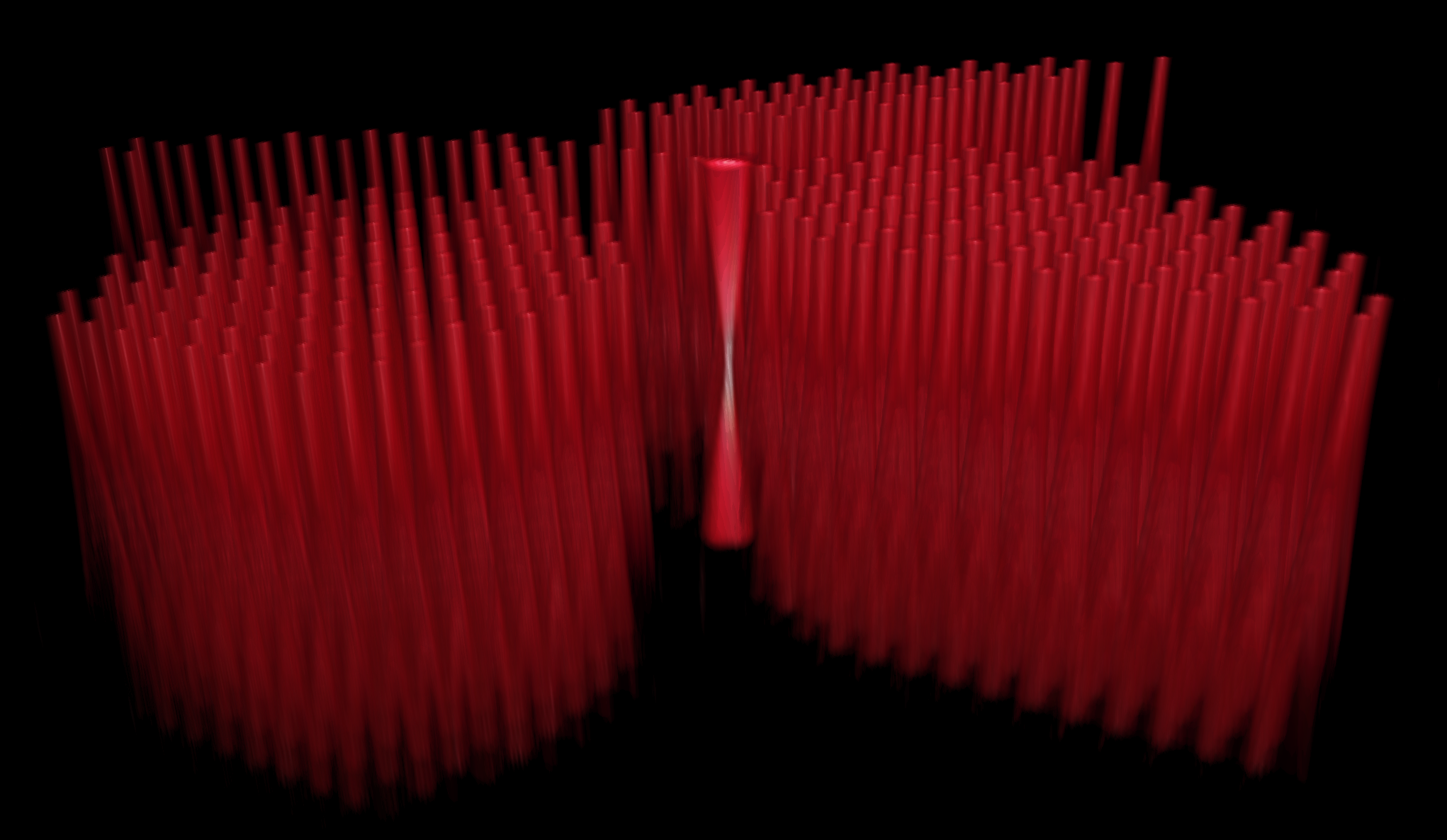}
    }
    \\
    \subcaptionbox{Sim, Bessel-like.}{
    	\includegraphics[height=7cm]{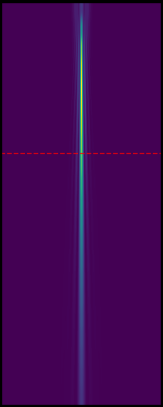}
    }
    \subcaptionbox{Experimentally captured reconstructions using Bessel-like beamshaper.}{
    	\includegraphics[height=7cm]{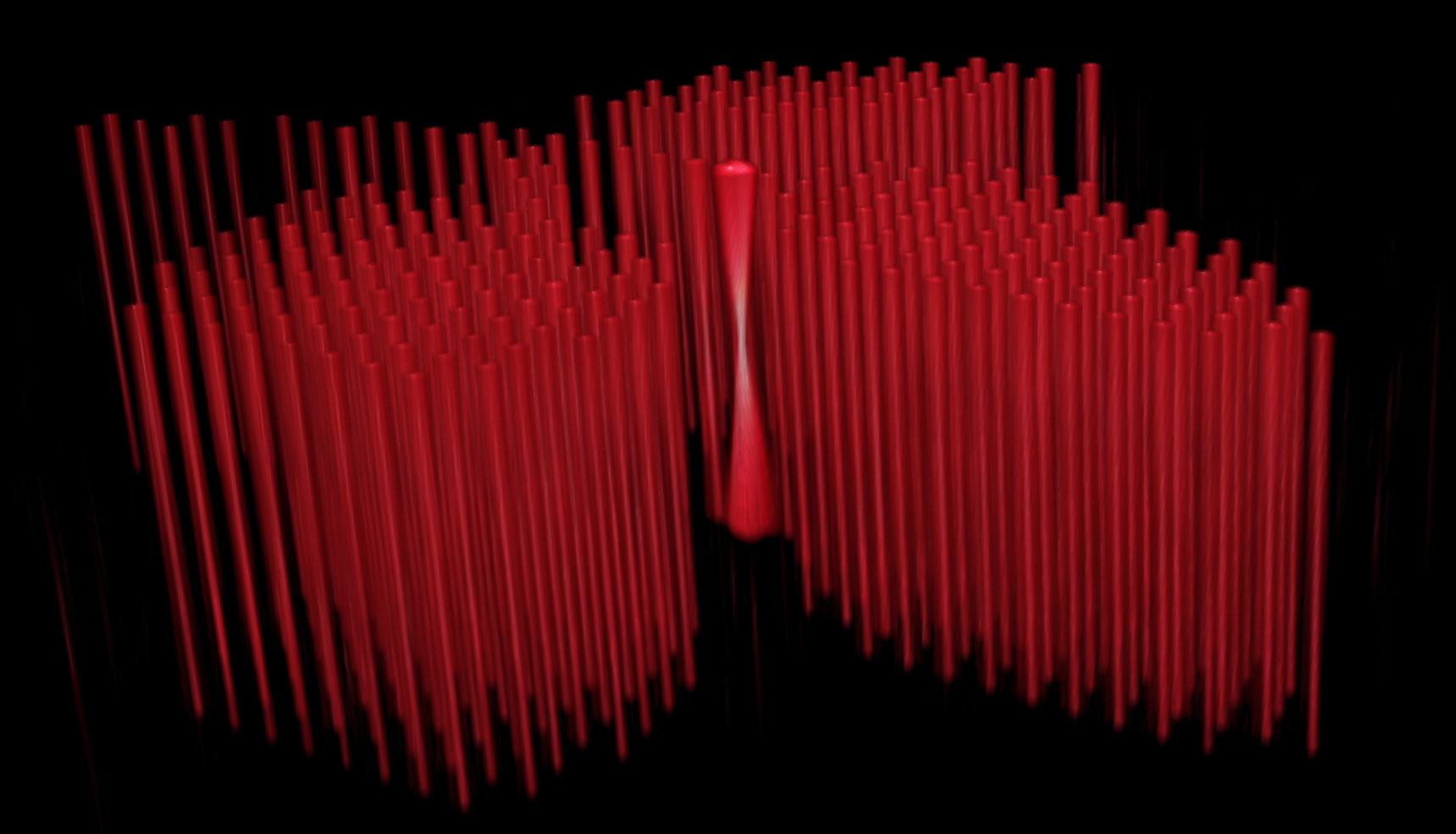}
    }
    \\
    \caption{Simulations (a, c, e) of the axial propagation of the beamshaping hologram around the focal plane (dashed red line) for the square, ring, and Bessel-like beam shaper, and experimentally (b, d, f) captured reconstructions through the focal plane using the same beamshapers. For both cases the propagation distance is 2 cm.}
    \label{fig:axial-propagation}
\end{figure*}

With HoloTile, the 3-dimensional PSF is altered according to the beamshaping hologram. Thus, the reconstruction in each output pixel evolves along the optical axis identically to each other output pixel. As is displayed in \cref{fig:axial-propagation}, the axial propagation of the beamshaping hologram is replicated in each output pixel in the experimental results. There is therefore clear control of the desired aligment between pixels, and how these may interact. Furthermore, axial beamshaping can also be engineered. For instance, by adding a quadratic phase term to the ring beam shaper, and thus translating the reconstruction plane axially, a Bessel-like beam (\cref{fig:axial-propagation}e,f) is generated through the focal plane. Comparing to the center zero-order, it is clear to see that the applicable axial extent of these output beams is greatly increased, and the intra-pixel interference is easily explained and predicted.

\section{HoloTile for Optogenetics}
Compared to conventional methods such as electrostimulation, MR or CT scanning, optogenetics offers a superior capability of stimulating specific types of neurons and observing their subsequent reactions. This allows for a comprehensive mapping of neuronal connections and interactions, which is essential for improving our understanding of brain functionality. Generally, opsins, which are photoresponsive microbial proteins that can be targeted to specific neurons, react to light in the visible spectrum, particularly blue light. By activating these opsins, responses of individual neurons can be observed, enabling activities such as mapping and understanding of underlying processes.
However, the application of visible light presents a challenge as it gets significantly scattered by brain tissue, resulting in loss of both spatial selectivity and signal intensity unless delivered very close to the target neurons. A viable solution is using near-infrared light for stimulation, as opsins can respond to this wavelength through two-photon processes. Near-infrared light has the advantage of lower scattering and absorption in the tissue, thus offering better penetration depth.

Two-photon processes are intensity-dependent and require high intensities to be noticeable, necessitating the use of laser pulses to prevent tissue heating and damage. This intensity-dependence can also be leveraged to provide spatial selectivity by concentrating the light at a specific point or plane, resulting in a higher rate of two-photon processes there compared to other points. To obtain useful currents for optogenetics, one must excite an extended region, preferably shaping the light pulse to match the shape of the neurons. The extremely efficient light-shaping of CGH presents numerous advantages for this particular application. Arbitrary dynamic intensity patterns can be generated and directed to selective locations in 3-dimensions with great precision. 

However, traditional CGH modalities suffer from both slow hologram generation \cite{madsen_holotile_2022,madsen_comparison_2022} and serious problems with speckle noise. Speckles are dramatically deteriorating, particularly in two-photon processes as is the case for near-infrared opsin excitation  due to the intensity squared dependence, $I^2$, and not only intensity, $I$, as is the case in standard one-photon excitation processes.

The tiling and subsequent PSF shaping of HoloTile allows for near-speckle-free, and extremely fast, generation of arbitrary intensity patterns. By using extremely high-resolution SLMs, perhaps in combination with existing light-sculpting modalities (see \cref{sec:gpc-holotile}), we anticipate the ability to address triple-digit interconnections of living neurons in a volume - without laser speckles. This may make it possible to fulfil the long-term vision of neuroscientists of creating true so-called ‘Circuit Optogenetics’, crucial for unravelling the cause of degenerative diseases such as Alzheimer’s or Parkinson’s and to better understand the complex nature of the human brain.

\section{HoloTile for Particle Manipulation}
Optical trapping techniques have typically made use of focused Gaussian beams (TEM00) in order to manipulate and examine high refractive index (RE) micro- and nanoscale particles. The particles are confined within the beam due to an equillibrium of counteracting forces on the particle i.e. gradient and scattering forces due to the high RE of the particle compared to the surrounding medium.
However, when attempting to trap low RE or strongly absorbing particles, this equilibrium is difficult, if not impossible, to reach. As such, the particles are actively repelled from regions of high light intensity. For these cases, dark optical traps have been realized \cite{daria_dynamic_2004}, consisting of dark regions surrounded by high-intensity light. The process of generating well-defined surrounded dark regions has been shown in a multitude of ways e.g. high-speed deflectable mirrors scanning a beam in a circular fashion or more complex beam shaping such as higher-order Laguerre-Gaussian beams.
While these methods work well for singular or a low number of traps, their associated mechanical complexity and implicit non-parallel nature do not aid in fully dynamic and independent manipulation of many particles. In \cref{fig:field-ring}, an experimental reconstruction using ring PSFs is shown. By employing a high numerical aperture (NA) objective lens, equivalent high-efficiency and high-contrast dynamic ring patterns can be synthesized for this exact application. Once the desired ring diameter is determined and the PSF shaping phase calculated, the sub-holograms of HoloTile can be re-calculated in real-time for precise parallel optical manipulation. CGH has long been used for particle manipulation and optical trapping in high-NA configurations \cite{grier_holographic_2006-1,daria_optical_2011,gluckstad_light_2017, suarez_optical_2021-1}. The active modulation of the 3D system PSF and the rapid hologram generation of HoloTile may aid in achieving even greater control in emerging fields such as light-robotics.

\section{HoloTile for Laser Material Processing}
\label{sec:lmp}
\begin{figure}
	\centering
	\includegraphics[width=\columnwidth]{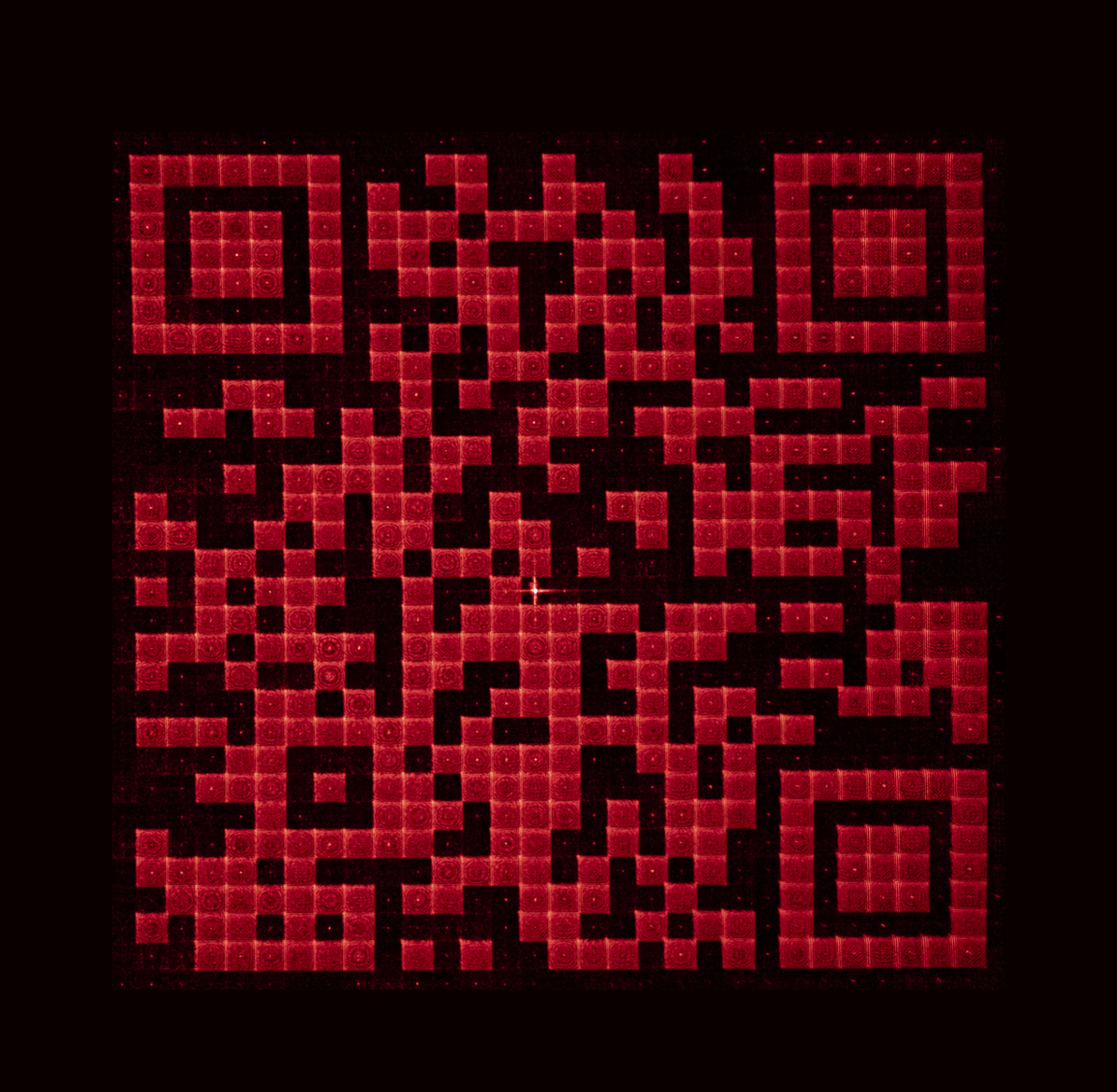}
	\caption{Experimental QR code reconstruction using the PSF shaped output pixels as individual pixels in the QR code.}
	\label{fig:qr}
\end{figure}

Laser beam shaping has found key applications in fields such as microbiology, neuroscience \cite{papagiakoumou_scanless_2010,papagiakoumou_optical_2013}, optical manipulation \cite{palima_wave-guided_2012}, materials processing and even consumer electronics.
Common light shaping techniques fall loosely into two groups; those shaping individual beams and those utilizing multiple shaped beams in parallel. In laser material processing (LMP), for applications such as engraving, welding, and machining, individual focused beams are typically used. These beams can be both simply focused laser beams with carefully selected Gaussian beam profiles, or more complex shapes e.g. rings, top-hats, and the like \cite{grunewald_influence_2021,bischoff_design_2019}. These individual beams can then be scanned laterally to facilitate a given process. While a lot can be accomplished with a single scanning beam, one-shot or parallel processing may open up the doors for faster throughput. The homogeneous patterning available in the reconstructions of HoloTile may allow for high-efficiency, high-fidelity, and dynamic material surface processing. For high-power applications, powerful pulsed lasers are typically used. While the spectral broadening associated with pulsed laser sources is typically not desirable in holography, it may be turned to an advantage in combination with temporal focusing (TF). 
In TF, the pulsed beam is chromatically separated in space by a diffraction grating. Focusing the dispersed beam in an objective lens forces the separated wavelengths to positively interfere only in its focal plane. The resulting reconstruction is an axially localized 2D reconstruction with high energy density. This may enable HoloTile-based holographic metal engraving or welding. One can easily imagine the usefulness of rapidly tagging (by engraving or similar process a pattern such as is shown in \cref{fig:qr}) an assembly line of products, without the need for additional products such as plastic stickers or ink printers.
 
\section{HoloTile for Volumetric 3D Printing}
\label{sec:vol-3d}
The industry and consumer standard for additive manufacturing for purposes such as rapid prototyping is still based on fused deposition modelling (FDM), in which a material is made pliable and extruded through a nozzle, and deposited layer-by-layer to form a final 3-dimensional object. While the simplicity of operation and speed of material deposition of FDM printers are desirable characteristics in many applications, for 3D models with increased fidelity and smaller feature sizes (smaller than the nozzle diameter), alternatives methods yield superior results. Stereolithographic printers (SLA) function by localized photopolymerization of a photosensitive resin under focused or patterned ultra-violet (UV) light. Common implementations of SLA printing feature a vat of UV-photosensitive resin containing a submerged and axially translatable build surface. Using a light transport method such as galvanically scanning a focused UV laser, or masked UV light-projection, each layer in a sliced 3D model can be cured in the resin. Following the curing of each layer, the build surface is moved mechanically to allow for the curing of the next layer at the appropriate height. While SLA printing can achieve impressive results in terms of fidelity, the printing speed is limited to the order of millimeters per hour (in terms of printing height) \cite{huang_review_2020}. In addition, both SLA and FDM printers typically require the printing of support struts in order to facilitate the fabrication of objects with complex hollow structures or overhangs.

Several techniques \cite{kelly_volumetric_2019,loterie_high-resolution_2020,regehly_xolography_2020,shusteff_one-step_2017,madridwolff_controlling_2022} have been proposed that address printing time, print fidelity, and the need for support struts.
Tomographic volumetric additive manufacturing (VAM) is a revolutionary 3D-printing approach where an entire three-dimensional object is simultaneously solidified by irradiating a resin or cell-filled hydrogel from multiple angles with dynamic light patterns in the ultraviolet wavelength regime. Tomographic VAM can print complex objects on the centimetre scale in a matter of seconds as opposed to hours without the need for supporting structures. Though tomographic VAM has the potential to produce highly complex structures with a higher throughput and a wider range of printable materials than the conventional layer-by-layer additive manufacturing of the aforementioned methods, the resolution is currently limited by the usually large \'etendue of the applied illumination system. Typically, a so-called Digital Light Projection (DLP) illumination system is applied based on light inefficient digital micro-mirror devices (DMD). The amount of light $I_v$ from the laser $I_0$ that is directed from the DMD into the 3D printing volume is given by
\begin{equation}
	I_v = \eta \frac{I_0\sum\limits_{i=0}^{N-1} \frac{p_i}{p_{max}}}{N}
\end{equation}
where $\eta$ accounts for inherent inefficiencies in the DMD (e.g., diffraction efficiency, deadspace, mirror reflectivity, etc), $p_i$ is the displayed gray value of the $i$'th pixel, $p_{max}$ is the maximum gray value of the DMD ($255$ for an 8-bit DMD), and $N$ is the number of pixels on the DMD. Thus, to achieve $100\%$ light-transport, all pixels on the DMD must be set to $p_{max}$. For the sparse tomographic projections usually calculated by the Radon transform, only a tiny fraction (in the order of  hundreds to thousands of pixels on megapixel DMDs) of the total micro-mirrors will be deflecting light towards the 3D bioprinting volume. Hence, a substantial light power source is inherently needed which typically implies a multimoded large \'etendue light source. 
We propose the inclusion of HoloTile into the VAM domain, as it exhibits certain qualities that may finally cement holography as not just a viable, but preferable, light-shaping modality for 3D-printing applications.

First, as with all holographic projection, the light-transport efficiency from laser to resin can, in the best case, approach $90-100\%$ (using high-efficiency, multiple-bit phase-only spatial light modulators (SLM)). Even binary phase holographic patterns (e.g. Lee holograms \cite{lee_iii_1978,conkey_high-speed_2012,rodrigo_high-speed_2006}) achieve theoretical diffraction efficency of $40\%$, vastly outperforming the light-transport of a DMD displaying the typical sparse Radon patterns that are typical for VAM. Using a Fourier holographic approach such as HoloTile utilizes the entire SLM for wavefront modulation. As such, even sparse target patterns may approach the much higher theoretical light-transport efficiency, thus allowing for rapid printing and much lower power requirements to the input laser device. Consequentially, a single spatial mode laser diode source with optimal light \'etendue can be used, thereby improving the spatial resolution of the 3D bioprinting.

Second, due to the controllability of the output pixels of the HoloTile reconstructions, these can be tailored to fit the 3D bioprinting application precisely. One may imagine that both extremely axially localized output pixels, and pixels with substantial axial extent (e.g. Bessel-Gauss or needle beams\cite{durnin_diffraction-free_1987,grunwald_thin_2007}) akin to what is shown in \cref{fig:axial-propagation}, will be preferential in different VAM configurations. The controllability also extends to time and volume. With the extremely fast hologram generation rate of HoloTile, arbitrary holograms can be multiplexed temporally and spatially in order to create more complex light-sculpting volumes, and thus higher fidelity 3D bioprinted objects.

Third, the rapid hologram generation rate may provide for real-time aberration-corrected VAM. Any refractive index mismatch between the resin or hydrogel, the cuvette, and the surrounding index matching liquid will be a cause of aberrations, and in turn, non-accurate replication of the target object. By taking advantage of the modular creation of the HoloTile holograms and the rapidity with which they are created, it may be possible to compensate for any possible aberration in the system in real-time. This possibility is elaborated in \cref{sec:abb-corr}.

Last, since we only require a collimated single-mode laser beam, SLM, and an optional Fourier lens to generate the HoloTile reconstructions, the mechanical complexity and cost of such a bioprinting setup can be reduced dramatically, hopefully allowing for true commercial use.

\begin{figure*}
	\centering
	\includegraphics[width=16cm]{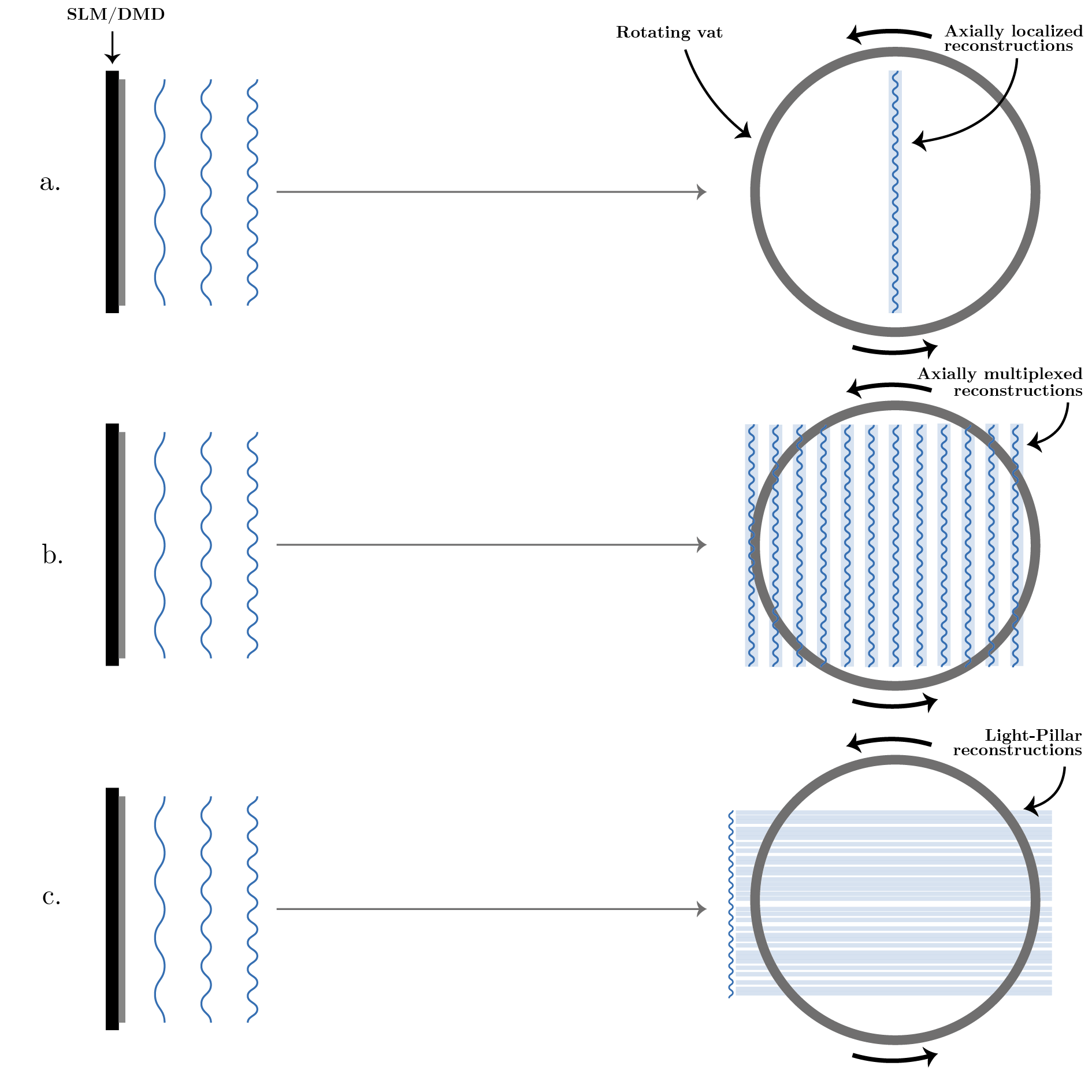}
	\caption{Three VAM modalities utilizing the properties of HoloTile. a) Axially localized reconstruction plane b) Axially multiplexed reconstruction planes form pseudo-``pixel-beams" c) Single hologram light pillars generated by the Bessel-like beamshaper form the ``pixel-beams".}
	\label{fig:VAMIll}
\end{figure*}

Several holographic VAM modalities present themselves when considering HoloTile as the light delivering engine. Considering the rotating resin vat approach \cite{loterie_high-resolution_2020-1}, we can speculate as to how the properties of HoloTile may be utilized optimally. We might imagine an instance of HoloTile employing some axial localization technique e.g., temporal focusing, in order to engineer the light such that the optical power is constrained only within the specified 2D reconstruction plane. Then, as the vat is rotated around this 2D plane, the light modulator can be encoded with easily calculable and angle specific holograms (\cref{fig:VAMIll}a).

Another instance may replicate the so-called tomographic ``pixel-beams" using either temporal multiplexing or specialized axial beamshaping. We might imagine that, for each projection angle in the tomographic printing process, a number of axially shifted \textit{identical} reconstruction planes can be constructed such as to mimic the overall effect of a z-extruded 3D pattern through the vat. For each vat rotation angle, the light modulator can, given a high enough framerate, sweep through the angle dependent axially shifted reconstructions (\cref{fig:VAMIll}b).

In a final instance, we consider the Bessel-like beams shown in \cref{fig:axial-propagation}f as a method for synthesizing the ``pixel-beams" using a single hologram per vat rotation angle. The Bessel-like beams will propagate through the cuvette and, due to the self-repairing property of Bessel beams \cite{fahrbach_microscopy_2010, zhao_propagation_2019}, perhaps be more resilient to scattering and diffraction by the curing resin.

Ultimately, we anticipate that our patented \cite{gluckstad_holographic_2022, gluckstad_holographic_2023} HoloTile light engine can pave the way for highly light efficient VAM of 3D bioprinted centimetre scale objects with optimal \'etendue and micron-sized features in a few tens of seconds. 
\begin{figure*}[h!]
	\centering
	\includegraphics{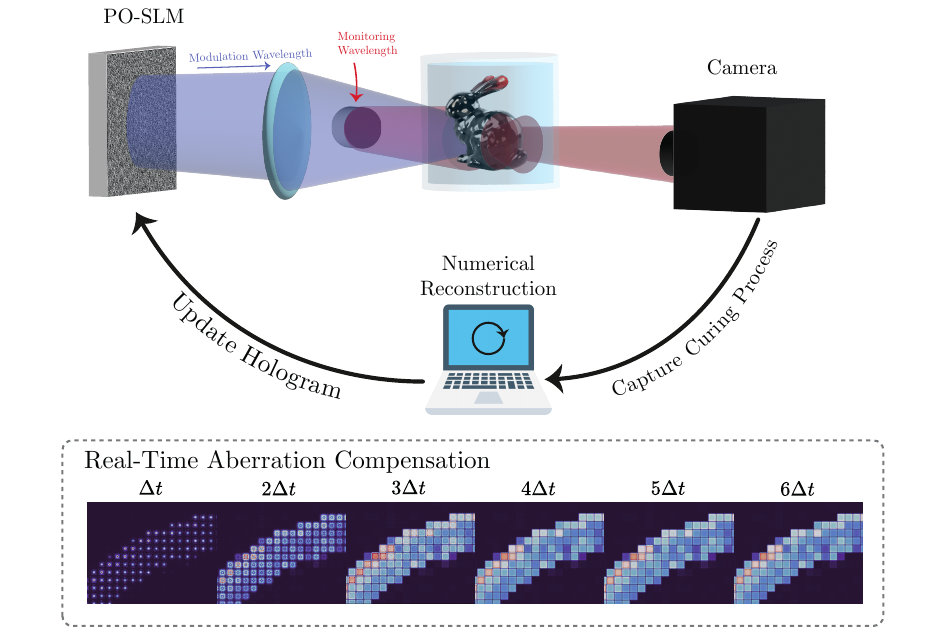}
	\caption{Abstracted aberration correction loop for VAM. As the reconstructions are projected into the printing volume, a camera monitors the progress. The rapid nature of HoloTile allows for fast corrections of the reconstructions.}
	\label{fig:aberration-correction-print}
\end{figure*}

\section{HoloTile for (Quantum) Information Transmission}
The helico-conical beams described in \cref{sec:hc}, and shown experimentally to work well in the HoloTile configuration in \cref{fig:field-hc}, are one instantiation of a class of beams denoted as optical vortices. Common for these beams are their field distribution dependence on a topological charge $\ell$ and their azimuthal angle $\theta$. A fundamental property of optical vortices is the inherent ability to carry orbital angular momentum (OAM) \cite{alonzo_helico-conical_2005,engay_interferometric_2019}.
Beams capable of carrying non-zero OAM e.g. Laguerre-Gaussian, higher-order Bessel, helico-conical, have been shown to dramatically increase data transfer rates \cite{wang_terabit_2012,shen_optical_2019} both in optical fibers and free-space communication, as they possess an additional degree of freedom that can be multiplexed. 

Another interesting property of OAM modes is the fact that they can be created in coherent superpositions, and can thus be exploited as \textit{qudits} in the area of quantum communication. A qudit can, as opposed to a binary qubit, be interpreted as a quantum particle with $d$ discrete states. The addition of multiple quantum states to a quantum computing, communication or encryption system not only allows for higher information density in transmission, but also for increased security, and more efficient computing \cite{erhard_twisted_2017,wootters_single_1982,bocharov_factoring_2017}.

Using HoloTile, we provide a method to rapidly generate numerous dynamic OAM beams in parallel, with excellent photon efficiency. One may imagine that the ability to address many such beams simultaneously, both for modulation, transmission relay, and read-out \cite{engay_interferometric_2019}, could, for an expert in the field, prove advantageous.

\section{GPC-Enhanced HoloTile}
\label{sec:gpc-holotile}
By utilizing a high-resolution phase-only SLM\cite{noauthor_gaea-2_2022}, which enables ultra high-efficiency reconstructions with excellent fidelity and phase control, any light-shaping that is performed is, naturally, limited to phase-only modulation. While this results in the required use of a phase-retrieval method such as the Gerchberg-Saxton algorithm\cite{gerchberg_practical_1972}, it also affects the PSF shaping that can be performed. For instance, the method with which each frequency component is shaped to be square is based on the stationary phase solution to the problem, given a phase-modulator that is laterally infinite\cite{dickey_gaussian_1996,dickey_laser_2014}. Since the aperture of a practical SLM is finite, the resulting reconstructions will deviate, however slightly, from the perfect target square \cite{madsen_new_2022}. This can indeed be seen in the reconstructions in \cref{fig:field-square}, where edges of the squares are softer than desired. Consequentially, the soft edges may also cause ``bleeding" into adjacent output pixels which, due to the unconstrained phases in the reconstruction plane, cause unwanted interference in the inter-pixel gaps. Now, if we lift the requirement of pure phase-modulation only from the PSF shaping step, we gain an additional degree of freedom in our ability to shape the beam into our desired output pixels.
A variety of light manipulation methods exist that can shape a common light source, generally a Gaussian laser beam, into basic shapes such as a rectangle or circle, all while preserving a continuous amplitude and phase profile. GPC, among the available phase-only alternatives, has proven to be an ideal substitute for amplitude modulation \cite{palima_gaussian_2008,gluckstad_optimal_2001,villangca_gpc-enhanced_2015,banas_holo-gpc_2017,palima_generalized_2007}. GPC, an enhancement of Zernike's award-winning phase contrast microscopy \cite{zernike_how_1955}, was primarily developed for beam shaping, but also applies to quantitative phase imaging \cite{palima_diffractive_2012}, optical encryption \cite{daria_phase-only_2004}, and emerging fields such as neuroscience \cite{papagiakoumou_scanless_2010} and atomtronics \cite{lee_analogs_2013}.
The light distributions generated by GPC in real-world applications mirror those seen with basic amplitude modulation. Both generate patterns with sharp outlines, continuous phase, and intensity. However, despite their qualitative similarities, they differ greatly in their photon efficiency. Amplitude modulation generates darkness by blocking or absorbing light, while GPC uses destructive and constructive interference to redirect photons to form the intended foreground pattern.

Thus, by utilizing a fixed and pre-optimized GPC shaped read-out beam, incident on the phase-only SLM, it may be possible to obtain much better defined output pixels, by tailoring not only the phase-modulation for PSF shaping, but also the amplitude, while maintaining the high diffraction efficiency.

\section{HoloTile Aberration Correction}
\label{sec:abb-corr}
While the HoloTile approach may offer lower speckle and more control, it is not impervious to inherent optical aberrations. Furthermore, the physical limitations of the optical setup introduces errors in the beam-shaping \cite{madsen_holotile_2022,madsen_new_2022}. However, because of the regularity and homogeneity of the output pixels and their propagation, and the simplicity with which they are controlled (analytical beam-shaping functions), these aberrations and beam-shaping errors may be more easily corrected in both an analytical and iterative manner. Since the output pixels are shaped by to the global PSF-shaping phase, they inherently describe the optical performance of the system, assuming that the phase function has been implemented correctly. This means that the optical system may be corrected continuously at a much more fundamental level than traditional CGH setups, since it obviates the need for an initial PSF characterization.

For example, as mentioned in \cref{sec:vol-3d}, when discussing volumetric 3D printing, additional hardware is necessary in order to negate optical effects of both the cuvette and resin to produce the appropriate curing patterns in the cuvette volume, i.e. the refractive index matching volume of mineral oil. It is not hard to imagine the unlikeliness of finding a liquid whose optical properties align perfectly with those of the resin and cuvette, especially considering that the cuvette and resin may not be an exact match either. Therefore, even granting a perfect aberration-free Fourier transform engine and SLM, slight differences in the optical properties of the cuvette assembly and printing volume will, inevitably, lead to unintended distortions of the projections. This problem is present in any application for which propagation through a refractive or scattering medium is required.

As such, any steps taken towards monitoring and correcting the wavefront of reconstructions will have immediate positive effects on all possible applications of HoloTile. In \cref{fig:aberration-correction-print}, the abstracted process of aberration correction in volumetric printing is shown. In this configuration, while a resin-curing light source is modulated by the SLM in order to form the 3D model, a monitoring light source at a non-curing wavelength is shone through the printing volume and onto a camera sensor. By capturing the printing progress, live corrections to dose amount may be employed in real-time such that the 3D volume is cured uniformly. Another configuration obviates the need for the additional monitoring light source by monitoring the reconstruction plane(s) of the modulated light source, and updating the holograms to ensure correct holograms and PSF shaping. Though aberration correction can be performed through iterative optimization, machine learning (ML) has also showed significant progress in the area.
Machine learning, and especially deep convolutional neural networks (CNN), have been used extensively for fast and accurate aberration correction\cite{saha_practical_2020,sharifzadeh_phase_2020,sharifzadeh_phase_2023,shi_end--end_2022,sirico_compensation_2022,tang_phase_2023,tian_dnn-based_2019,wang_correction_2020,wang_deep_2021} in optical sensing, imaging, and projection. Even for machine learning methods wherein aberration correction is not the explicit goal, the end-to-end design of certain network architectures ensures implicit correction, as long as these corrections are either constant between training and inference, or general enough during training to be useful for correcting arbitrary aberrations. For instance, several strides have been made in ML-aided holographic reconstruction \cite{huang_holographic_2021,huang_phase_2020,huang_phase_2021,rivenson_phase_2018,ren_end--end_2019,wang_eholonet_2018,wang_y-net_2019,sinha_lensless_2017,ju_learning-based_2022,madsen_-axis_2023} in which either raw holograms or back-propagated holograms are processed in custom trained networks to generate more accurate, aberration-reduced, reconstructions.

\section{Conclusion}
HoloTile presents new opportunities for CGH to impact research and industry. With the original paper and patent \cite{gluckstad_holographic_2022,madsen_holotile_2022}, we showed the process of reducing unwanted nearest-neighbor speckle noise, stemming from overlapping PSFs and undetermined phases, by shaping the individual frequency components in Fourier holography. Unlike the original paper, we have here attempted to highlight the ability for HoloTile to generate high-resolution reconstructions, in addition to the PSF shaping. We also show HoloTile in a lensless setting, further reducing the mechanical complexity.

 New PSF shaping phase profiles, including disks, rings, lines and optical vortices have been shown, broadening the possible areas where HoloTile may be advantageous. We have introduced possible applications such as laser material processing and volumetric additive manufacturing, where the high-efficiency, high-speed advantages of HoloTile can aid in reducing processing times. In a combination with temporal focusing, high-power pulsed lasers may be used to selectively and dynamically engrave or weld metals using HoloTile reconstructions.
 In optogenetics and particle trapping and manipulation, HoloTile enables the use of dynamic and parallel beam-shaping to rapidly address hundreds, if not thousands, of e.g. cells and particles with high light-efficiency. And due to the decreased speckle noise, unwanted non-linear excitations can be reduced drastically. The output pixels can be tuned in shape and size for precise and selective addressing.
Optical vortices and their associated orbital angular momentum have found uses in exceptionally fast tele-communication and quantum information transfer. HoloTile provides a solution to manipulate a multitude of beams with non-zero orbital angular momentum in parallel.

\ack
This work has been supported by the Novo Nordisk Foundation, Denmark (Grand Challenge Program; \\ NNF16OC0021948) and the Innovation Fund Denmark.

\section*{References}
\bibliographystyle{iopart-num}
\bibliography{references_update}

\end{document}